\documentclass[conference]{IEEEtran}
\IEEEoverridecommandlockouts
\usepackage{cite}
\usepackage{amsmath,amssymb,amsfonts}
\usepackage{algorithmic}
\usepackage{graphicx}
\usepackage{textcomp}
\usepackage{subcaption}
\usepackage{enumitem}
\usepackage{xcolor}
\usepackage{tikz}
\usepackage{xspace}
\usepackage{makecell}
\usepackage{tabularx}
\usepackage{booktabs}
\usepackage{balance}
\usepackage{dblfloatfix}    

\usetikzlibrary{positioning,fit,calc,shapes.geometric}
\usepackage{soul} 
\usepackage{hhline}
\definecolor{lightyellow}{RGB}{250, 250, 180}
\definecolor{HLYELLOW}{RGB}{240, 127, 0}
\definecolor{hlyellow}{RGB}{240, 127, 0}
\sethlcolor{lightyellow}
\definecolor{lightcyan}{RGB}{160,255,255}

\definecolor{r1}{RGB}{87,114,158}
\definecolor{r2}{RGB}{204,137,99}
\definecolor{r3}{RGB}{196,78,82}
\definecolor{r4}{RGB}{93,157,107}

\usepackage[textwidth=0.7cm,colorinlistoftodos]{todonotes}
\usepackage{etoolbox}

\newtoggle{usetodo}
\toggletrue{usetodo} 

\usepackage{mdframed}
\global\mdfdefinestyle{review}{%
linecolor=lightyellow,linewidth=3pt,%
leftmargin=0cm,rightmargin=0cm,%
skipabove=0cm,skipbelow=0cm,%
innerrightmargin=0cm,innerleftmargin=0cm,%
innerbottommargin=0cm,innertopmargin=0cm,%
backgroundcolor=lightyellow
}

\usepackage{marginnote}

\def\BibTeX{{\rm B\kern-.05em{\sc i\kern-.025em b}\kern-.08em
    T\kern-.1667em\lower.7ex\hbox{E}\kern-.125emX}}

\usepackage{multirow}
\usepackage[switch]{lineno}
\usepackage{url}
\usepackage{fancyhdr} 
\usepackage{array}
\newcolumntype{P}[1]{>{\centering\arraybackslash}p{#1}}
\newcolumntype{M}[1]{>{\centering\arraybackslash}m{#1}}
\usepackage[strict]{changepage}
\usepackage[normalem]{ulem} 
\newcommand{\dani}[1]{[\hl{#1}]$\mbox{}_\mathtt{daniele}$}
\newcommand{\salvo}[1]{[\hl{#1}]$\mbox{}_\mathtt{salvo}$}
\newcommand{\htor}[1]{[\hl{#1}]$\mbox{}_\mathtt{htor}$}

\newcommand{\slingshot}[0]{\textsc{Slingshot}\xspace}
\newcommand{\aries}[0]{\textsc{Aries}\xspace}
\newcommand{\rosetta}[0]{\textsc{Rosetta}\xspace}

\newcommand{\crystal}[0]{\textsc{Crystal}\xspace}
\newcommand{\malbec}[0]{\textsc{Malbec}\xspace}
\newcommand{\shandy}[0]{\textsc{Shandy}\xspace}

\begin{document}

\DeclareDocumentCommand\review{m g g}{%
    {\IfNoValueF {#2}{%
    \IfNoValueF {#3}{%
    {\marginnote{\sethlcolor{#3}\hl{\normalfont \textbf{{\normalsize{\color{white}#2}}}}}%
    }%
    }%
    \IfNoValueT {#3}{%
    {\marginnote{\normalfont \textbf{\normalsize{#2}}}%
    }%
    }%
    }%
    \hl{#1}%
    }%
}

\title{An In-Depth Analysis of the Slingshot Interconnect
}

\makeatletter
\newcommand{\linebreakand}{%
  \end{@IEEEauthorhalign}
  \hfill\mbox{}\par
  \mbox{}\hfill\begin{@IEEEauthorhalign}
}
\makeatother

\author{\IEEEauthorblockN{Daniele De Sensi}
\IEEEauthorblockA{\textit{Department of Computer Science} \\
\textit{ETH Zurich}\\
ddesensi@ethz.ch}
\and
\IEEEauthorblockN{Salvatore Di Girolamo}
\IEEEauthorblockA{\textit{Department of Computer Science} \\
\textit{ETH Zurich}\\
salvatore.digirolamo@inf.ethz.ch}
\and
\IEEEauthorblockN{Kim H. McMahon}
\IEEEauthorblockA{\textit{Hewlett Packard Enterprise}\\
kim.mcmahon@hpe.com}
\linebreakand
\IEEEauthorblockN{Duncan Roweth}
\IEEEauthorblockA{\textit{Hewlett Packard Enterprise}\\
duncan.roweth@hpe.com}
\and
\IEEEauthorblockN{Torsten Hoefler}
\IEEEauthorblockA{\textit{Department of Computer Science} \\
\textit{ETH Zurich}\\
torsten.hoefler@inf.ethz.ch}
}

\maketitle

\begin{abstract}
The interconnect is one of the most critical components in large scale computing systems, and its impact on the performance of applications is going to increase with the system size. In this paper, we will describe \slingshot, an interconnection network for large scale computing systems. \slingshot is based on high-radix switches, which allow building exascale and hyperscale datacenters networks with at most three switch-to-switch hops. Moreover, \slingshot provides efficient adaptive routing and congestion control algorithms, and highly tunable traffic classes. \slingshot uses an optimized Ethernet protocol, which allows it to be interoperable with standard Ethernet devices while providing high performance to HPC applications. We analyze the extent to which \slingshot provides these features, evaluating it on microbenchmarks and on several applications from the datacenter and AI worlds, as well as on HPC applications. We find that applications running on \slingshot are less affected by congestion compared to previous generation networks.
\end{abstract}


\section{Introduction}
The first US exascale supercomputer will be built within two years, marking an important milestone for computing systems. Exascale computing has been a long-awaited goal, which required significant contributions both from academic and industrial research. One of the most critical components having a direct impact on the performance of such large systems is the interconnection network (\textit{interconnect}). Indeed, by analyzing the performance of the Top500 supercomputers~\cite{top500} when executing HPL~\cite{linpack} and HPCG~\cite{hpcg}, two benchmarks commonly used to assess supercomputing systems, we can observe that HPCG is typically characterized by $\sim$50x lower performance compared to HPL. Part of this performance drop is caused by the higher communication intensity of HPCG, clearly showing that, among others, an efficient interconnection network can help in exploiting the full computational power of the system. The impact of the interconnect on the performance of supercomputing systems increases with the scale of the system, highlighting the need for novel and efficient solutions.

Both the HPC and datacenter communities are following a path towards convergence of HPC, data centers, and AI/ML workloads, which poses new challenges and requires new solutions. Workloads are becoming much more data-centric, and large amounts of data need to be exchanged with the outside world. Due to the wide adoption of Ethernet in datacenters, interconnection networks should be compatible with standard Ethernet, so that they can be efficiently integrated with standard devices and storage systems. Moreover, many data center workloads are latency-sensitive. For such applications, \textit{tail latency} is much more relevant than the best case or average latency. For example, web search nodes must provide $99^{th}$ percentile latencies of a few milliseconds~\cite{tailscale}. This is also a relevant problem for HPC applications, whose performance may strongly depend on messages latency, especially when using many global or small messages synchronizations. Despite the efforts in improving the performance of interconnection networks, tail latency still severely affect large HPC and data center systems~\cite{appawarerouting,gpcnet,tailscale,10.1145/3302424.3303973}.

To address these issues, Cray\footnote{Cray is a \textit{Hewlett Packard Enterprise} (HPE) company since 2019.} recently designed the \slingshot interconnection network. \slingshot will power all three announced US exascale systems~\cite{aurora,frontier,elcapitan} and numerous supercomputers that will be deployed soon. It provides some key features, like adaptive routing and congestion control, that make it a good solution for HPC systems but also for cloud data centers. 
\slingshot switches have 64 ports with 200 Gb/s each and support arbitrary network topologies. To reduce tail latencies, \slingshot offers advanced adaptive routing, congestion control, and quality of service (QoS) features. Those also protect applications from interference, sometimes referred to as network noise~\cite{appawarerouting,htornnoise}, caused by other applications sharing the interconnect.
Lastly, \slingshot brings HPC features to Ethernet, such as low latency, low packet overhead, and optimized congestion control, while maintaining industry standards. In \slingshot, each port of the switch can negotiate the available Ethernet features with the attached devices, and can communicate with existing Ethernet devices using standard Ethernet protocols, or with other \slingshot switches and NICs by using \slingshot specific additions. This allows the network to be fully interoperable with existing Ethernet equipment while at the same time providing good performance for HPC systems.

In this study, we experimentally analyze \slingshot's performance features to guide researchers, developers, and system administrators. We use Mellanox ConnectX-5 100 Gb/s Ethernet NICs to test the ability of \slingshot to deal with standard \textit{RDMA over Converged Ethernet} (RoCE) traffic\footnote{200 Gb/s Ethernet NICs were not available at the time of writing}. Moreover, by doing so we can analyze the impact of the switch on the end-to-end performance by factoring out some of the improvements on the Ethernet protocol introduced by \slingshot. We first analyze the latencies of a quiet system. Then, we analyze the impact of congestion on both microbenchmarks and real applications for different configurations, showing that \slingshot is only marginally affected by \textit{network noise}. To further show the benefits of the congestion control algorithm, we compare \slingshot to Cray's previous \aries network, which has a similar topology and uses a similar routing algorithm.

\section{Slingshot Architecture}\label{sec:slingshot}
We now describe the \slingshot interconnection network. We first introduce the \rosetta switch and show how switches can be connected to form a \textit{Dragonfly}~\cite{dragonfly} topology. We then dive into specific features of \slingshot such as adaptive routing, congestion control, and quality of service management. Lastly, we describe the main characteristics of the \slingshot additions to Ethernet and the software stack.

\subsection{Switch Technology (\rosetta)}\label{sec:slingshot:rosetta}
The core of the \slingshot interconnect is the \rosetta switch, providing 64 ports at 200 Gb/s per direction. Each port uses four lanes of 56 Gb/s \textit{Serializer/Deserializer} (SerDes) blocks using \textit{Pulse-Amplitude Modulation} (PAM-4) modulation. Due to \textit{Forward Error Correction} (FEC) overhead, 50Gb/s can be pushed through each lane. The \rosetta ASIC consumes up to 250 Watts and is implemented on TSMC’s 16 nm process. \rosetta is composed by 32 peripheral function blocks and 32 tile blocks. The peripheral blocks implement the \textit{SerDes}, \textit{Medium Access Control} (MAC), \textit{Physical Coding Sublayer} (PCS), \textit{Link Layer Reliability} (LLR), and Ethernet lookup functions. 

The 32 tile blocks implement the crossbar switching between the 64 ports, but also adaptive routing and congestion management functionalities. The tiles are arranged in four rows of eight tiles, with two switch ports handled per tile, as shown in Figure~\ref{fig:rosetta-tiles}. 
The tiles on the same row are connected through 16 per-row buses, whereas the tiles on the same column are connected through dedicated channels with per-tile crossbars. Each row bus is used to send data from the corresponding port to the other 16 ports on the row. 
The per-tile crossbar has 16 inputs (i.e., from the 16 ports on the row) and 8 outputs (i.e., to the 8 ports on the column). For each port, a multiplexer is used to select one of the four inputs (this is not explicitly shown in the figure for the sake of clarity).
Packets are routed to the destination tile
through two hops maximum. 
Figure~\ref{fig:rosetta-tiles} shows an example: if a packet is received on \textit{Port 19} and must be routed to \textit{Port 56}, the packet is first routed on the row bus, then it goes through the 16-to-8 crossbar highlighted in the picture, and then down a column channel to \textit{Port 56}. Thanks to the hierarchical structure of the tiles, there is no need for a 64 ports arbiter, and the packets only incur in a 16 to 8 arbitration. 

\begin{figure}[h]
    \centering
    \includegraphics[width=\columnwidth]{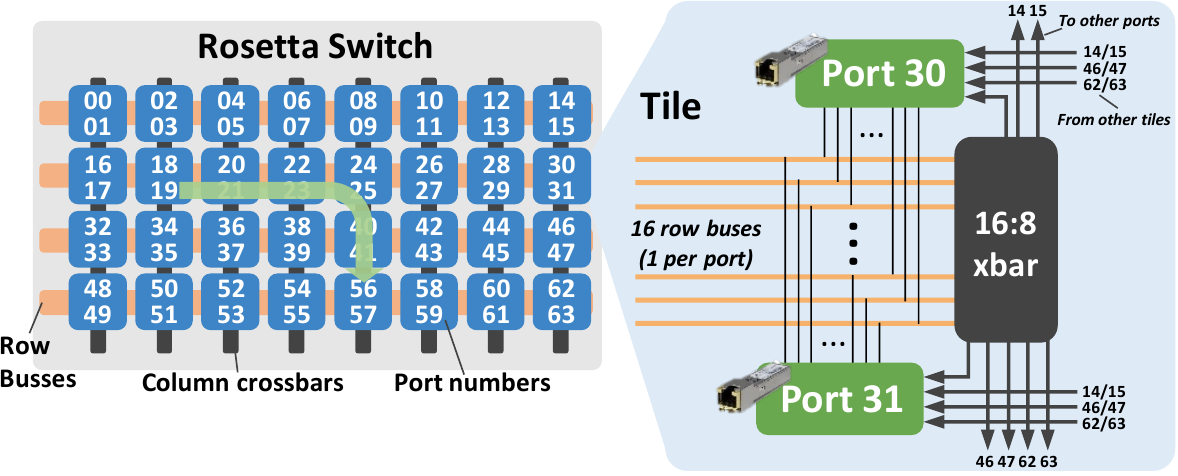}
    \caption{\rosetta switch tiled structure.}
    \label{fig:rosetta-tiles}
\end{figure}

The 32 tiles in \rosetta implement a crossbar between the 64 ports. For performance and implementation reasons, the crossbar is physically composed by different function-specific crossbars, each handling a different aspect of the switching traffic:
\begin{itemize}
\item \textbf{Requests to Transmit} To avoid \textit{head-of-line blocking} (HOL)~\cite{hol}, \rosetta relies on a virtual output-queued architecture~\cite{voq1, voq2} where the routing path is determined before sending the data. The data is buffered in the input buffers until the resources are available, guaranteeing no further blocks. Before forwarding the data, a request-to-transmit is sent to the tile corresponding to the switch output port. When a \textit{grant to transmit} is received from the output port, the data is forwarded. 
\item \textbf{Grants to Transmit} Grants to transmit are sent by the tile handling the output port to the tile from which the switch received the packet. In the previous example, the grants would be transmitted from the tile handling \textit{Port 56}, to the tile handling \textit{Port 19}. Grants are used to notify the permission to forward the data to the next hop. The use of requests and grants to transmit is a central piece of the QoS management. 
\item \textbf{Data} Data is sent on a wider crossbar (48B). To speed up the processing, \rosetta parses and processes the packet header as soon as it arrives, even if the data might still be arriving. 
\item\textbf{Request Queue Credits} Credits provide an estimation of queue occupancy. This information is then used by the adaptive routing algorithm (see Section~\ref{sec:slingshot:routing}) to estimate the congestion of different paths and to select the least congested one.
\item \textbf{End-to-End Acks} End-to-End acknowledgments are used to track the outstanding packets between every pair of network endpoints. This information is used by the congestion control protocol (see Section~\ref{sec:slingshot:congestion}). 
\end{itemize}

By using physically separated crossbars, \slingshot guarantees that different types of messages do not interfere with each other and that, for example, large data transfers do not slow down requests and grants to transmit.

\begin{figure}[htpb]
    \centering
    \includegraphics[width=\columnwidth]{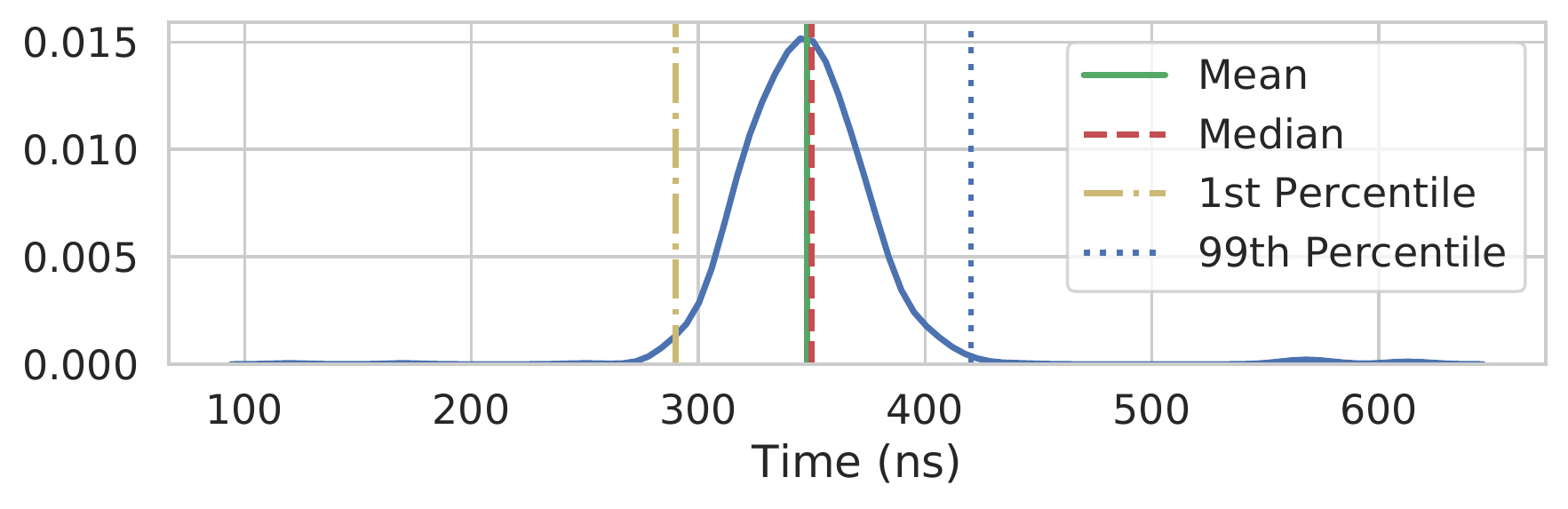}
    \caption{Distribution of switch latency for RoCE traffic.}
    \label{fig:switch-latency}
\end{figure}

To analyze the impact of the switch architecture on the latency, we report in Figure~\ref{fig:switch-latency} the latency of the switch when dealing with RoCE traffic. It is worth remarking that, because we are using standard RoCE NICs, the NIC sends plain Ethernet frames, and we cannot exploit all the features of \slingshot's specialized Ethernet protocol (Section~\ref{sec:slingshot:ethernet}). Some of the features like \textit{link-level reliability} and propagation of congestion information are however still used in the switch-to-switch communications.
To compute the latency of the switch, we consider the latency difference between 2-hops and 1-hop latencies (we provide details on the topology in  Section~\ref{sec:slingshot:topology}). We observe that \rosetta has a mean and median latency of 350 nanoseconds, with all the distribution lying between 300 and 400 nanoseconds, except for a few outliers.

\subsection{Topology}\label{sec:slingshot:topology}
\rosetta switches can be arranged into any arbitrary topology. \textit{Dragonfly}~\cite{dragonfly} is the default topology for \slingshot-based systems, and it is the topology we refer to in the rest of the paper. Dragonfly is a hierarchical direct topology, where all the switches are connected to both computing nodes and other switches. Sets of switches are connected between each other forming so-called \textit{groups}. The switches inside each group may be connected by using an arbitrary topology, and groups are connected in a fully connected graph. In the \slingshot implementation of Dragonfly (shown in Figure~\ref{fig:topology}), each \rosetta switch is connected to 16 endpoints through copper cables (up to 2.6 meters), using the remaining 48 ports for inter-switches connectivity. The partitioning of these 48 ports between inter- and intra-group connectivity, as well as the number of switches per group, depends on the size of the system. In \slingshot, the switches inside a group are always fully connected through copper cables. Switches in different groups are connected through long optical cables (up to 100 meters). Due to the full-connectivity both within the group and between groups, this topology has a diameter of 3 switch-to-switch hops. 
\begin{figure}[h]
    \centering
    \includegraphics[width=\columnwidth]{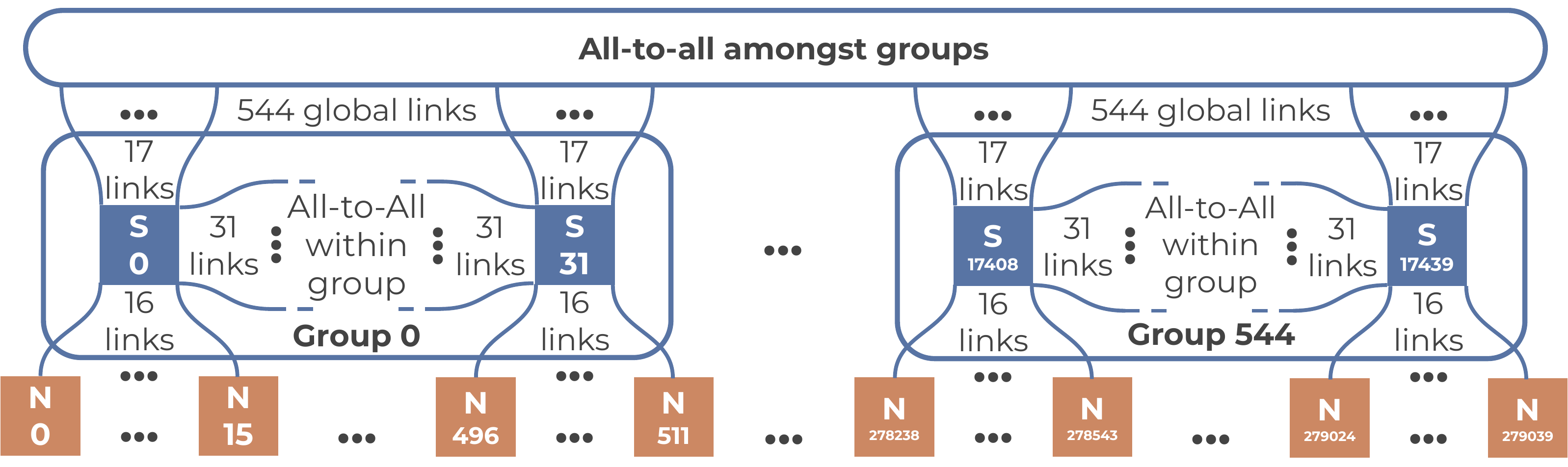}
    \caption{\slingshot Topology. In this specific example we show the topology of the largest 1-dimensional Dragonfly network that can be built with the 64-ports \rosetta switches.}
    \label{fig:topology}
\end{figure}

\begin{figure}[htpb]
  \centering
  \includegraphics[width=\linewidth]{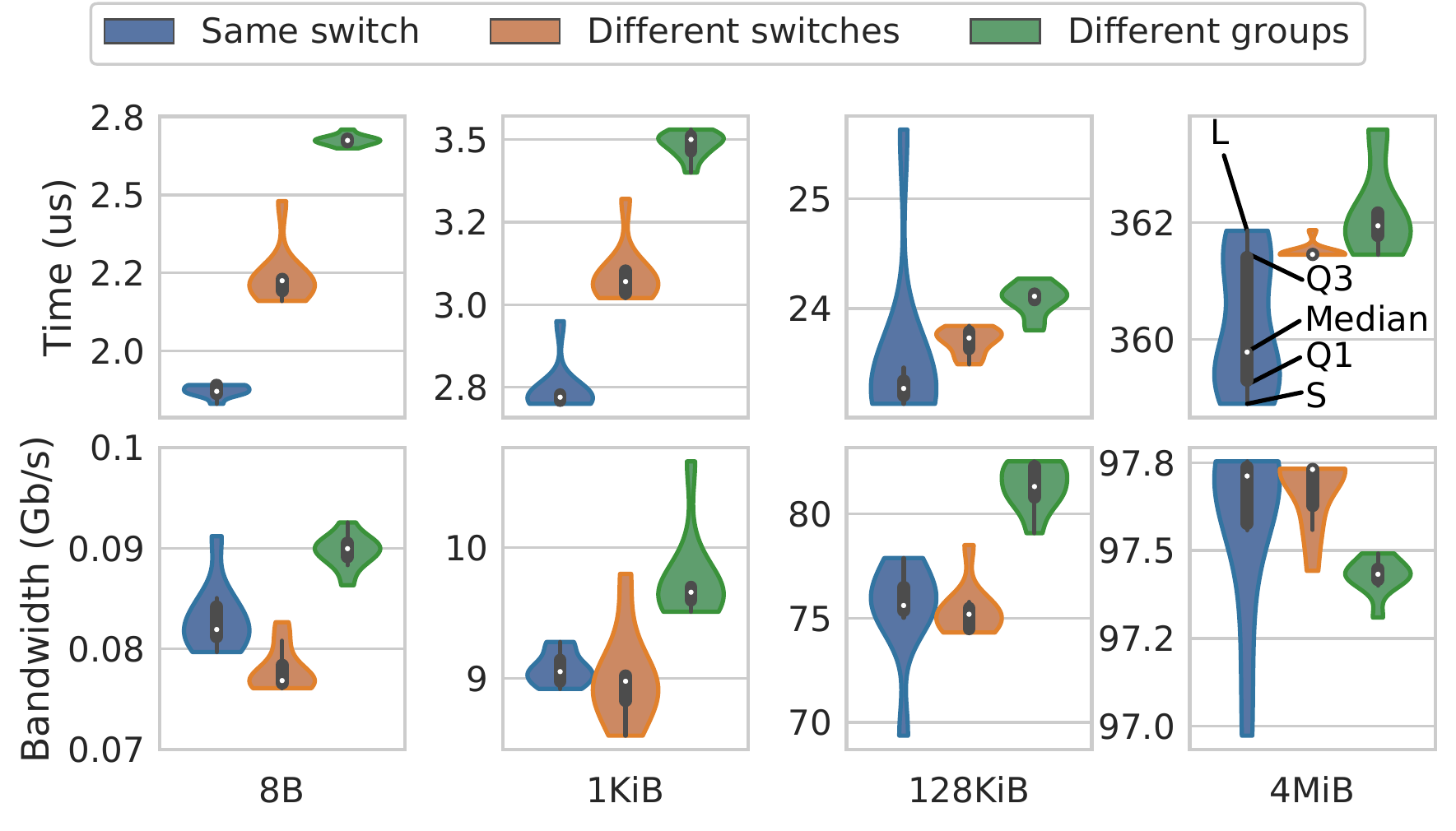}
\caption{Latency and bandwidth for different node distances, on an isolated system. $Q1$ is the first quartile, $Q3$ is the third quartile, $IQR = Q3 - Q1$, $S$ is the smallest sample greater than $Q1 - 1.5 \cdot IQR$, and $L$ is the largest sample smaller than $Q3 + 1.5 \cdot IQR$.}
\label{fig:evaluation:distance}
\end{figure}

Thanks to the low-diameter, applications performance only marginally depend on the specific node allocation. We report in Figure~\ref{fig:evaluation:distance} the latency and the bandwidth between nodes at different distances, and for different message sizes on an isolated system. We consider nodes connected to ports on the \textit{same switch} (1 inter-switch hop), connected to two different switches in the \textit{same group} (2 inter-switch hops), and connected to two \textit{different switches} in two different groups (3 inter-switch hops). For the \textit{same switch} case, we observed no significant difference when using two ports on the same switch tile or on two different tiles. 

First, we observe that, in the worst case, the node allocation has only a $40\%$ impact on the latency for $8B$ messages and that, starting from $16KiB$ messages we observe less than $10\%$ difference in latency between the different node distances. The same holds for bandwidth, with less than $15\%$ difference between the different distances across all the message sizes. In some cases, we observe a slightly higher bandwidth when the nodes are in two different groups, because more paths connect the two nodes, increasing the available bandwidth.

In the largest system (shown in Figure~\ref{fig:topology}), each group has 32 switches (for a total of $32 \times 16 = 512$ nodes, and switches inside each group are fully connected by using 31 switches ports. The remaining 17 ports from each switch are used to globally connect all the groups in a fully connected network. In this specific case, because each group contains 32 switches and each switch uses 17 ports to connect to other groups, each group has $32 \times 17 = 544$ connection towards other groups. This leads to a system having 545 groups, each of which is connected to 512 nodes, for a total of $279\,040$ endpoints at full global bandwidth\footnote{In practice, the addressing scheme limits the number of groups to 511, for a total of $261\,632$ nodes.}. This number of endpoints satisfies both exascale supercomputers and hyperscale data centers demand. Indeed, this is larger than the number of servers used in data centers~\cite{amzcld}, and much larger than the number of nodes used by Summit~\cite{summit}, the most performing supercomputer at the time being, that currently relies on $4\,608$ nodes and delivers $200PFlop/s$. Thanks to this large number of endpoints, each computing node can have multiple connections to the same network, increasing the injection bandwidth and improving network resiliency in case of NICs failures.

\subsection{Routing}\label{sec:slingshot:routing}
In Dragonfly networks (including \slingshot), any pair of nodes is connected by multiple \textit{minimal} and \textit{non-minimal} paths~\cite{dragonfly,fatpaths}. For example, by considering the topology in Figure~\ref{fig:topology}, the minimal path connecting \textit{N0} to \textit{N496} includes the switches \textit{S0} and \textit{S31}. In smaller networks, due to links redundancy,multiple minimal paths are connecting any pair of nodes~\cite{fatpaths}. On the other hand, a possible non-minimal path involves an intermediate switch that is directly connected to both \textit{S0} and \textit{S31}. The same holds for nodes located in different groups. In this case, a non-minimal path crosses an intermediate group.

Sending data on minimal paths is clearly the best choice on a quiet network. However, in a congested network, with multiple active jobs, those paths may be slower than longer but less congested ones. To provide the highest throughput and lowest latency, \slingshot implements adaptive routing: before sending a packet, the source switch estimates the load of up to four minimal and non-minimal paths and sends the packet on the best path, that is selected by considering both the paths' congestion and length. The congestion is estimated by considering the total depth of the request queues of each output port. This congestion information is distributed on the chip by using a ring to all the forwarding blocks of each input port. It is also communicated between neighboring switches by carrying it in the acknowledgement packets. The total overhead for congestion and load information is an average of four bytes in the reverse direction for every packet in the forward direction.
As more packets take non-minimal paths and therefore average hop count per packet increases, both the latency and the link utilization increase. Therefore, \slingshot adaptive routing biases packets to take minimal paths more frequently, to compensate for the higher cost of non-minimal paths.

\subsection{Congestion Control}\label{sec:slingshot:congestion}
Two types of congestion might affect an interconnection network: \textit{endpoint} congestion, and \textit{intermediate} congestion~\cite{gpcnet}. The endpoint congestion mostly occurs on the last-hop switches, whereas intermediate congestion is spread across the network. Adaptive routing improves network utilization and application performance by changing the path of the packets to avoid intermediate congestion. However, even if adaptive routing can bypass congested intermediate switches, all the paths between two nodes are affected in the same way by endpoint congestion. As we show in Section~\ref{sec:evaluation:congestion}, this was a relevant issue on other networks, particularly for \textit{many-to-one} traffic. In this case, due to the highly congested links on the receiver side, the adaptive routing would spread the packets over the different paths but without being able to avoid congestion, because it is occurring in the last hop. 

Congestion control helps in mitigating this problem by decreasing the injection bandwidth of the nodes generating the congestion. However, existing congestion control mechanisms (like \textit{ECN}~\cite{ecn} and \textit{QCN}~\cite{qcn,dcqcn}) are not suited for HPC scenarios. They work by marking packets that experience congestion. When a node receives a packet that has been marked, it asks the sender to slow down its injection rate. These congestion control algorithms work relatively well in presence of large volume and stable communications (known as \textit{elephant flows}), but tend to be fragile, hard to tune~\cite{ecnordelay, dcqcnplus}, and generally unsuitable for bursty HPC workloads. Indeed, in standard congestion control algorithms, the control loop is too long to adapt fast enough, and while converging to the correct transmission rate, the offending traffic can still interfere with other applications.


To mitigate this problem, \slingshot introduces a sophisticated congestion control algorithm, entirely implemented in hardware, that tracks every in-flight packet between every pair of endpoints in the system. \slingshot can distinguish between jobs that are victims of congestion and those who are contributing to congestion, applying stiff and fast backpressure to the sources that are contributing to congestion. By tracking all the endpoints pairs individually, \slingshot only throttles those streams of packets who are contributing to the endpoint congestion, without negatively affecting other jobs or other streams of packets within the same job who are not contributing to congestion.
This frees up buffers space for the other jobs, avoiding HOL blocking across the entire network, and reducing tail latencies, which are particularly relevant for applications characterized by global synchronizations. 

The approach to congestion control adopted by \slingshot is fundamentally different from more traditional approaches such as ECN-based congestion control~\cite{ecn,qcn}, and leads to good performance isolation between different applications, as we show in Section~\ref{sec:evaluation:congestion}. 

\subsection{Quality of Service (QoS)}\label{sec:slingshot:qos} 
Whereas congestion control partially protects jobs from mutual interference, jobs can still interfere with each other. To provide complete isolation, in \slingshot jobs can be assigned to different traffic classes, with guaranteed quality of service. QoS and congestion control are orthogonal concepts. Indeed, because traffic classes are expensive resources requiring large amounts of switch buffers space, each traffic class is typically shared among several applications, and congestion control still needs to be applied within a traffic class.

Each traffic class is highly tunable and can be customized by the system administrator in terms of priority, packets ordering required, minimum bandwidth guarantees, maximum bandwidth constraint, lossiness, and routing bias~\cite{appawarerouting}. The system administrator guarantees that the sum of the minimum bandwidth requirements of the different traffic classes does not exceed the available bandwidth. Network traffic can be assigned to traffic classes on a per-packet basis.
The job scheduler will assign to each job a small number of traffic classes, and the user can then select on which class to send its application traffic. In the case of MPI, this is done by specifying the traffic class identifier in an environment variable. Moreover, communication libraries could even change traffic classes at a per-message (or per-packet) granularity.
For example, MPI could assign different collective operations to different traffic classes. For example, it may assign latency-sensitive collective operations such as \texttt{MPI\_Barrier} and \texttt{MPI\_Allreduce} to high-priority and low-bandwidth traffic classes, and bulk point-to-point operations to higher bandwidth and lower priority classes.

Traffic classes are completely implemented in the switch hardware. A switch determines the traffic class required for a specific packet by using the \textit{Differentiated Services Code Point} (DSCP) tag in the packet header~\cite{dscp}. Based on the value of the tag, the switch assigns the packet to one of the multiple virtual queues. Each switch will allocate enough buffers to each traffic class to achieve the desired bandwidth, whereas the remaining buffers will be dynamically allocated to the traffic which is not assigned to any specific traffic class.

\subsection{Ethernet Enhancements}\label{sec:slingshot:ethernet}
To improve interoperability, and to better suit datacenters scenarios, \slingshot is fully Ethernet compatible, and can seamlessly be connected to third-party Ethernet-based devices and networks. \slingshot provides additional features on top of standard Ethernet, improving its performance and making it more suitable for HPC workloads. \slingshot uses this enhanced protocol for internal traffic, but it can mix it with standard Ethernet traffic on all ports at packet-level granularity. This allows \slingshot to achieve high-performance, while at the same time being able to communicate with standard Ethernet devices, allowing it to be used efficiently in both supercomputing and datacenter worlds. 

To improve performance, \slingshot reduces the 64 Bytes minimum frame size to 32 Bytes, allows IP packets to be sent without an Ethernet header, and removes the inter-packet gap.
Lastly, \slingshot provides resiliency at different levels by implementing low-latency \textit{Forward Error Correction} (FEC)~\cite{fec}, \textit{Link-Level Reliability} (LLR) to tolerate transient errors, and lanes degrade~\cite{ldegrade} to tolerate hard failures. Moreover, the \slingshot NIC provides end-to-end retry to protect against packet loss. These are relevant features in high-performance networks. For example, FEC is required for all Ethernet systems at 100Gb/s or higher, independently from the system size, and LLR is useful in large systems (such as hyperscale data centers) to localize the error handling and reduce end-to-end retransmission. 

\subsection{Software Stack}

Communication libraries can either use the standard TCP/IP stack or, in case of high-performance communication libraries such as \textit{MPI}~\cite{mpi,mpi-exascale}, \textit{Chapel}~\cite{10.5555/2875144}, \textit{PGAS}~\cite{pgas} and \textit{SHMEM}~\cite{shmem}, the \textit{libfabric} interface~\cite{libfabric}. 
Cray contributed with new features to the \textit{libfabric} open-source verbs provider and \textit{RxM} utility provider to support the \slingshot hardware. All HPC traffic is layered over \textit{RDMA over Converged Ethernet} (RoCEv2) and data is sent over the network through packets containing up to 4KiB of data plus headers and trailers. Headers and trailers include Ethernet (26 bytes including the preamble), IPv4 (20 bytes), UDP (8 bytes), InfiniBand (14 bytes), and an additional RoCEv2 CRC (4 bytes), for a total of 62 bytes. 
\textit{Cray MPI} is derived from \textit{MPICH}~\cite{mpich} and implements the MPI-3.1 standard.  Proprietary optimizations and other enhancements have been added to \textit{Cray MPI} targeted specifically for the \slingshot hardware. Any MPI implementation supporting \textit{libfabric} can be used out of the box on \slingshot. Moreover, standard API for some features, like traffic classes, have been recently added to \textit{libfabric} and could be exploited as well. We report in Figure~\ref{fig:software-stack} the latencies for different message sizes and for different network protocols. We observe that for small message sizes, MPI adds only a marginal overhead to \textit{libfabric}.

\begin{figure}[h]
\centering
\begin{subfigure}{.68\linewidth}
    \centering
    \includegraphics[width=\columnwidth]{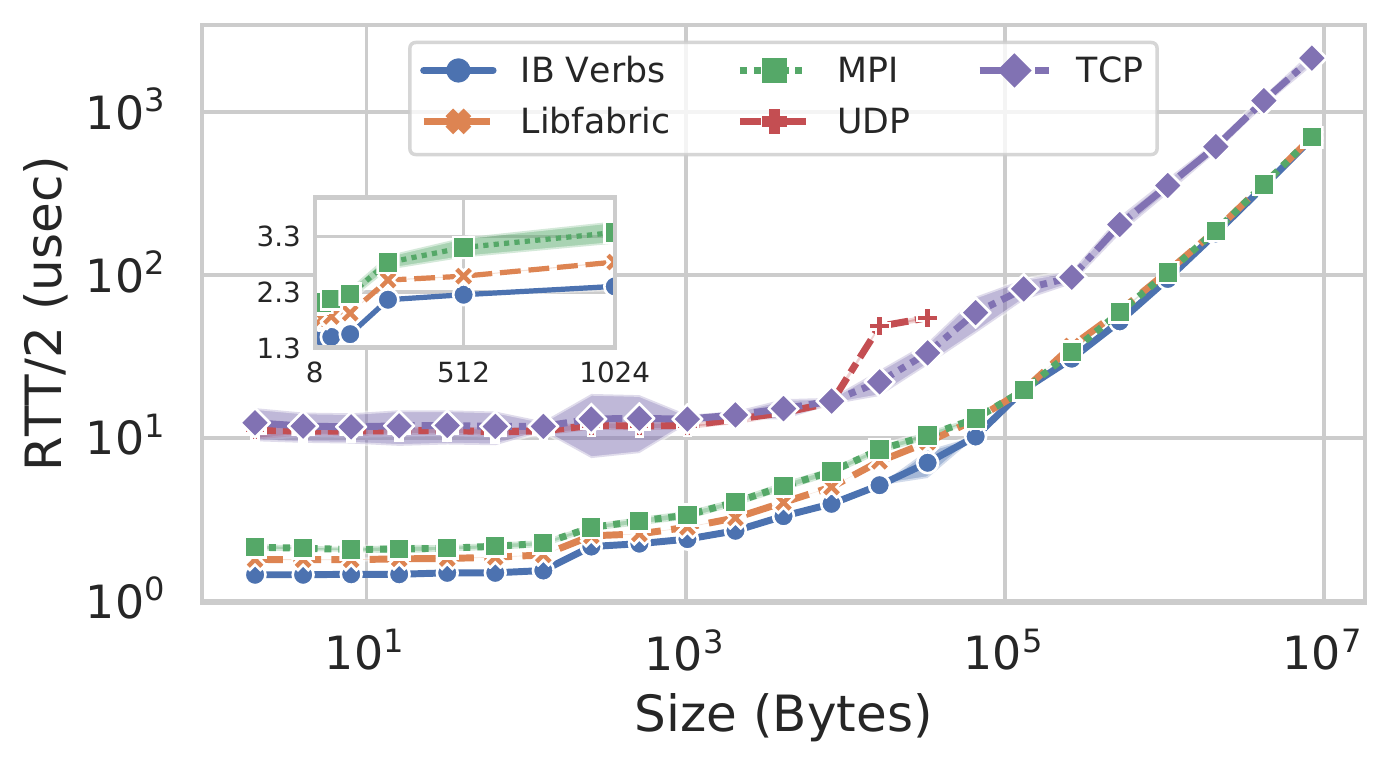}
\end{subfigure}
\hfill
\begin{subfigure}{.30\linewidth}
    \vspace{-1.5em}
    \centering
    \includegraphics[width=\columnwidth]{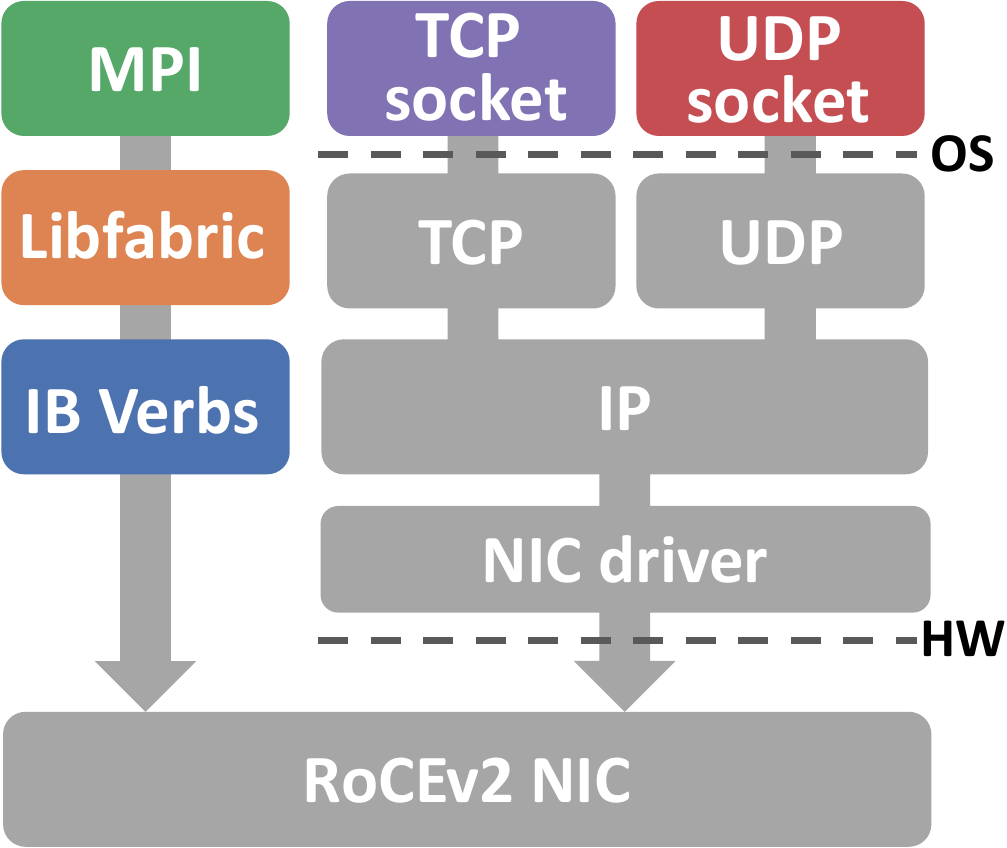}
\end{subfigure}
\caption{Half round trip time (RTT/2) for different message sizes (x-axis) and software layers.}
\label{fig:software-stack}
\end{figure}

Moreover, we show in Figure~\ref{fig:alltoall} the bisection bandwidth (i.e., the bandwidth when half of the nodes send data to the other half of the nodes and vice versa) and the \texttt{MPI\_Alltoall} bandwidth on \shandy, a \slingshot-based system using $1\,024$ nodes (see Section~\ref{sec:evaluation} for details). We report the results for different processes per node (PPN) and different message sizes. This system is composed of eight groups, and all the bisection cuts cross the same number of links. In this system, each group has 56 global links out of 112 (8 towards each other group), to match the injection bandwidth. Each of the 4 groups in one partition is connected to each of the 4 groups in the other partition, and the total number of links crossing a bisection cut is $4\cdot4\cdot8 = 128$. Because each link has a 200Gb/s bandwidth, and we are sending traffic in both directions, the peak bisection bandwidth is $128\cdot 200Gb/s \cdot 2 = 6.4Tb/s$. 

\begin{figure}[h]
    \centering
    \includegraphics[width=\columnwidth]{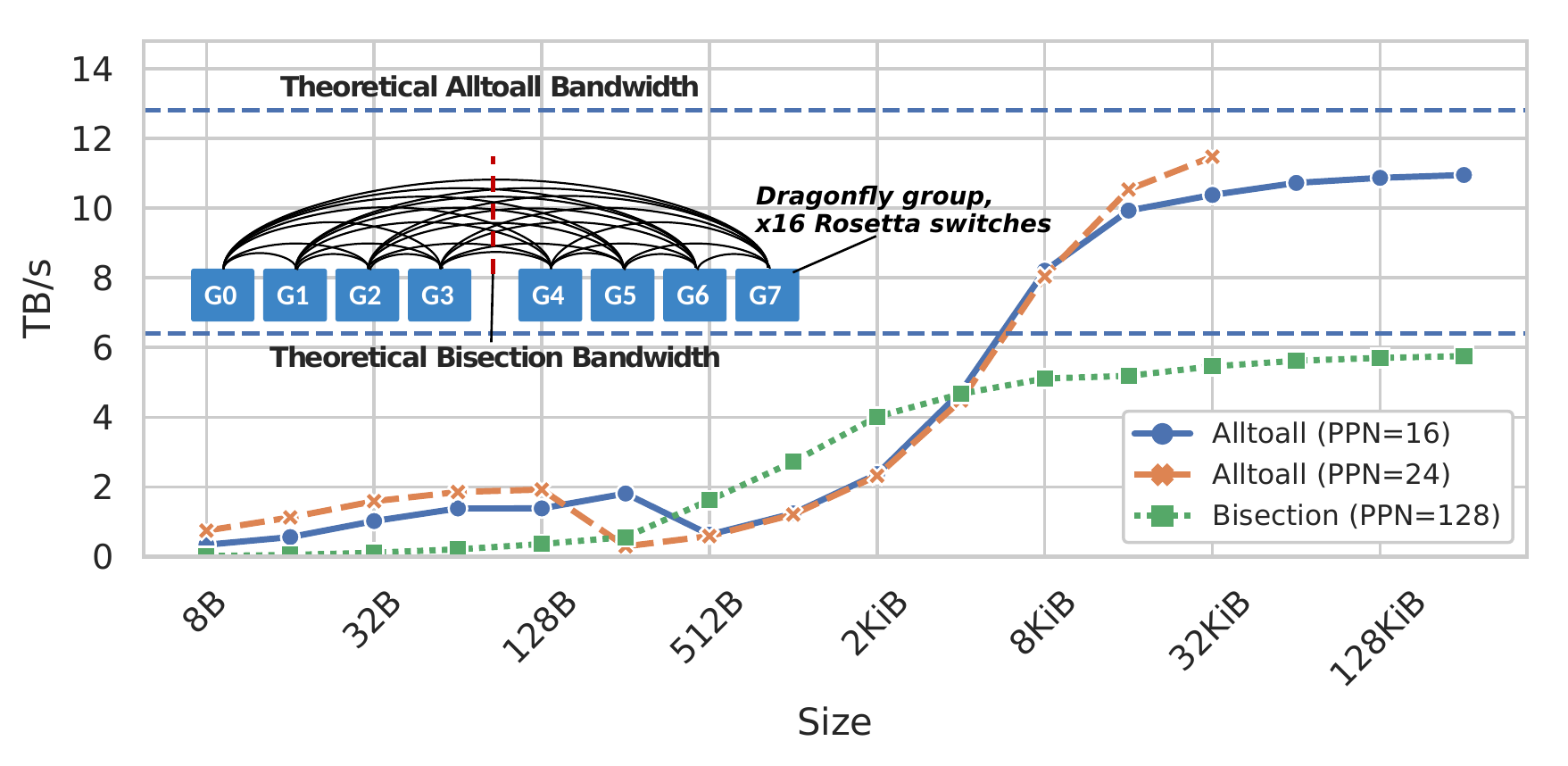}
    \caption{Bisection and \texttt{MPI\_Alltoall} bandwidth on all the $1\,024$ nodes of \shandy, for different \textit{processes per node} (PPN) and message sizes. The x-axis is in logarithmic scale.}
    \label{fig:alltoall}
\end{figure}

In an \textit{all-to-all} communication, each node sends $7/8$ of the traffic to nodes in the other 7 groups and $1/8$ of the traffic to nodes in the same group. Because this system has $56 \cdot 8 = 448$ global links, the \textit{all-to-all} maximum bandwidth is $8/7 \cdot 448\cdot 200Gb/s = 12.8Tb/s$.
Note that \texttt{MPI\_Alltoall} can achieve twice the bisection bandwidth because half of the connections terminate in the same partition~\cite{prisacari-ics-bandwidth-opt-alltoall}. The plot shows that the \texttt{MPI\_Alltoall} reaches more than the 90\% of the theoretical peak bandwidth, without any packet loss.
We observe a performance drop for 256 bytes messages because, to reduce memory usage, the MPI implementation switches to a different algorithm~\cite{optcoll} for messages larger than 256 bytes.

\begin{table*}[htpb]
\footnotesize
\begin{center}
\begin{tabularx}{\textwidth}{p{0.8cm}m{1.2cm}p{15cm}}
\textbf{\textsc{Type}}  & \makecell[c]{\textbf{\textsc{Appl.}}} & \makecell{\centering \textbf{\textsc{Description}}} \\ 
\toprule
\multirow{3}{*}{
\makecell{\\\\\vspace{-1em}\textsc{\textbf{HPC}}\\\\\includegraphics[width=0.04\textwidth]{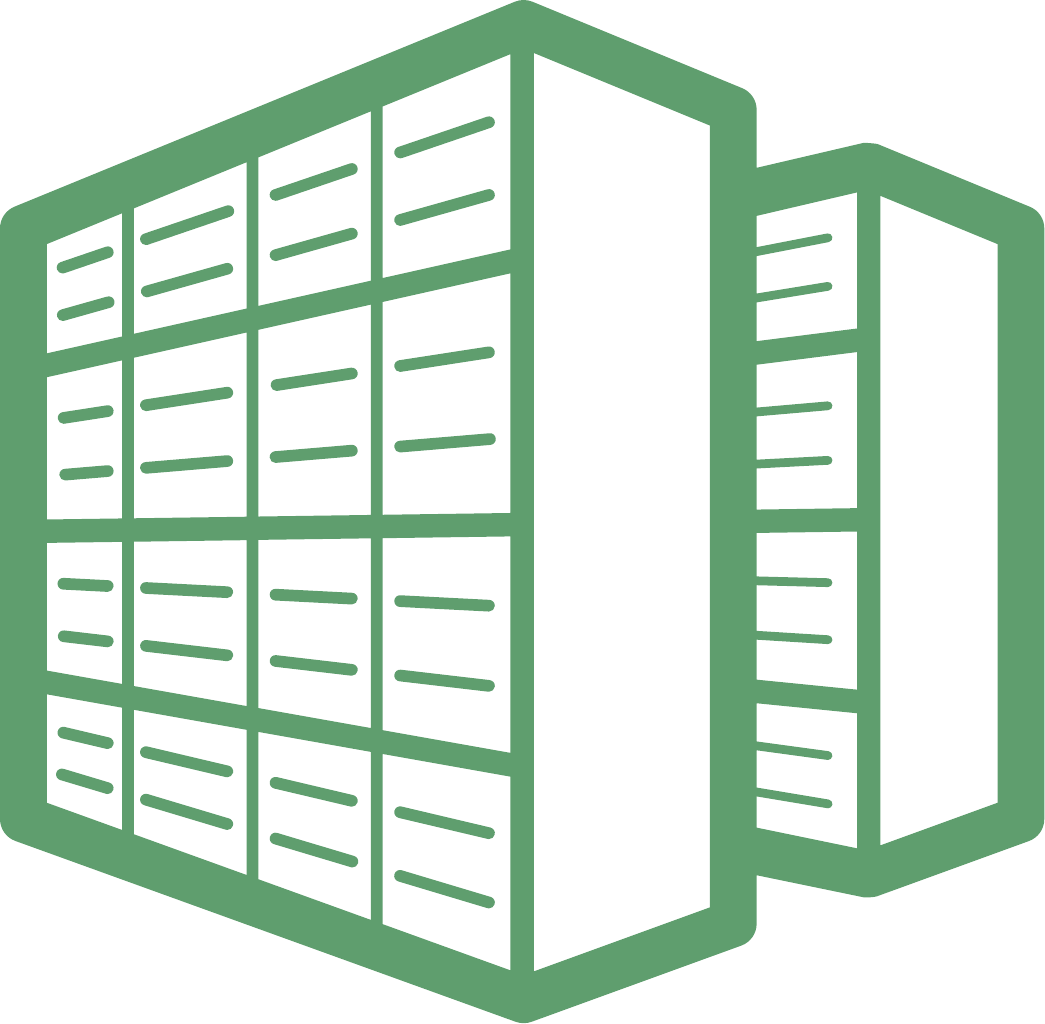}}}  & \makecell[c]{MILC} & \makecell[l]{It is a set of numerical simulation codes working on quantum chromodynamics (QCD)~\cite{milc}. We use the \texttt{su3\_rmd} kernel, that \\ decomposes a four dimensional grid, and mostly performs point-to-point neighbour communications and global reductions~\cite{milc-modeling}.} \\ 
\cmidrule{3-3}  & 
 \makecell[c]{HPCG} & \makecell[l]{A set of communication and computational patterns matching a wide set of applications. It relies on  sparse triangular solvers \\ and preconditioned conjugate gradient algorithms~\cite{hpcg}. It mostly uses stencil communications and global reductions.}\\ 
 \cmidrule{3-3} & 
 \makecell[c]{LAMMPS} & \makecell[l]{A molecular dynamics code that models an ensemble of particles in a liquid, solid, or gaseous state~\cite{lammps}. This kernel performs \\ reductions and point-to-point blocking and non-blocking communications, between nodes at different distances.} \\ 
 \cmidrule{3-3} & 
 \makecell[c]{FFT} & \makecell[l]{\textit{Fast Fourier Transform} on a 3D domain~\cite{fftw}. It employs broadcasts, scatters, and point-to-point communications~\cite{fftw:chara}.}\\ 
 \midrule \midrule & 
 \makecell[tc]{Resnet-\\proxy} & \makecell[tl]{This is a ML/AI proxy application~\cite{resnet}, reproducing the communication phases of a Deep500 benchmark~\cite{deep500} \\ \textit{Residual Neural Network} (resnet). This application uses non-blocking reduction operations.}  \\ 
\cmidrule{3-3}
\multirow{2}{*}{\makecell{\\\vspace{-1em}\textsc{\textbf{DC}}\\\\\includegraphics[width=0.04\textwidth]{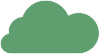}}} & 
\makecell{Silo} & \makecell[l]{A fast in-memory transactional database~\cite{silo}. Widely used in online transaction processing systems (OLTP).} \\  
\cmidrule{3-3} & 
\makecell{Sphinx} & \makecell[l]{A speech recognition system~\cite{Walker04sphinx-4:a}, involving probabilistically pruning a large search tree.} \\ 
\cmidrule{3-3} & 
\makecell{Xapian} & \makecell[l]{A search engine~\cite{xapian} using a search index built from a snapshot of the English version of Wikipedia. Multiple queries are \\ executed, with a distribution similar to that of online search queries.} \\ 
\cmidrule{3-3} & 
\makecell{Img-dnn} & \makecell[l]{An application using a deep neural network-based autoencoder to identify handwritten characters~\cite{imgdnn}.}  \\ 
\bottomrule
\end{tabularx}
\end{center}
\caption{Applications used as victim in the congestion tests. We consider both HPC and datacenter (DC) applications. Img-dnn, Xapian, Sphinx and Silo are all single-client, single-server applications, coming from the \textit{Tailbench} benchmark~\cite{tailbench} for latency-sensitive datacenter applications. We selected this subset because it covers a wide range of latencies, from microseconds (\textit{Silo}) to seconds (\textit{Sphinx}).}
\label{tab:apps}
\end{table*}

\section{Performance Study}\label{sec:evaluation}
We now study the performance of the \slingshot interconnect on real applications and microbenchmarks, by focusing on two key features of \slingshot, namely congestion control and quality of service management.
%
%
For our analysis, we consider the following systems:
\begin{itemize}
\item \crystal: A system based on the Cray \aries interconnect~\cite{crayxcpdf}. This system has 698 nodes. The CPUs on the nodes are \textit{Intel Xeon E5-269x}. The system is composed of two groups, each containing at most 384 nodes.
\item \malbec: A \slingshot system with 484 nodes. CPUs on the nodes are either \textit{Intel Xeon Gold 61xx} or  \textit{Intel Xeon Platinum 81xx} CPUs. The system is composed of four groups, each containing at most 128 nodes. Each group is connected to each other group through 48 global links operating at 200Gb/s each. Each node has a \textit{Mellanox ConnectX-5 EN} NIC. 
\item \shandy: A \slingshot system with 1024 nodes. Compute nodes are equipped with \textit{AMD EPYC Rome} 64 cores CPUs. The system is composed of eight groups, each containing 128 nodes. Each group is connected to each other group through 56 global links operating at 200Gb/s each. Each node has two \textit{Mellanox ConnectX-5 EN} NICs, each connected to a different switch of the same network, allowing a better load distribution and resilience in the event of NICs failures. 
\end{itemize}
We consider two \slingshot systems, of different size, to analyze the performance at different system scales. For all the experiments, we booked these systems for exclusive use, to have a controlled environment and avoid interference caused by other users.

\subsection{Congestion Control}\label{sec:evaluation:congestion}

To evaluate the ability of \slingshot to react to congestion, we divide the nodes in the system in two partitions: \textit{victim} nodes and \textit{aggressor} nodes. The aggressor nodes generate congestion that impacts the performance of victim nodes. We consider two types of congestion patterns: \textit{endpoint} congestion and \textit{intermediate} congestion, and we use the GPCNet code~\cite{gpcnet} to generate those congestion patterns. We generate endpoint congestion through a \textit{many-to-one} (\textit{incast}) communication pattern, where a number of nodes send data to the same endpoint by using \texttt{MPI\_Put}, and \textit{intermediate} congestion by using an \textit{all-to-all} pattern implemented through \textit{MPI\_Sendrecv}. Both aggressors exchange 128KiB messages. This decision is based on characterization studies on production systems, that show an average message size of $\sim 10^5$ bytes both in collective and point-to-point communications~\cite{mpi_characterization}. 

We consider the victim applications described in Table~\ref{tab:apps}. Moreover, we also analyze the impact of congestion on microbenchmarks, include standard MPI operations, and the \textit{ember} microbenchmarks~\cite{ember} reproducing some common communication patterns in HPC applications (\textit{halo3d}, \textit{sweep3d}, and \textit{incast}).
We first consider the results on 512 nodes. Then, we show the results for different node counts. We consider the following victim/aggressor splits: 460/52 ($\sim 90\%/10\%$), 256/256 ($\sim 50\%/50\%)$ and  53/459 ($\sim 10\%/90\%$). 
Because the implementation of some MPI collectives changes according to the number of nodes used, we have chosen these splits so that we run the victim with both power of two (256), even (460) and odd (53) number of nodes.
To further increase the generated congestion, in some experiments we increase the number of processes per node (\textit{PPN}) used by the aggressor. Each node used by the aggressor spawns PPN processes, each of them performing the same communications. Namely, the congestion pattern is concurrently executed PPN times.

Moreover, the allocation of the nodes to victims and aggressors determines how many switches and groups are shared between the two jobs and has a direct impact on the performance of the victim. In our experiments, we consider the three well-known allocation placement strategies~\cite{prisacari-dragonfly-mapping} depicted in Figure~\ref{fig:allocations}: \textit{linear}, where we allocate the first $n$ nodes to the victim and the remaining nodes to the aggressor; \textit{interleaved}, where we interleave the nodes allocated to the victim and the aggressor; and \textit{random}, where we randomly allocate the nodes to the victim and the aggressor. 


\begin{figure}[!b]
    \centering
    \includegraphics[width=\columnwidth]{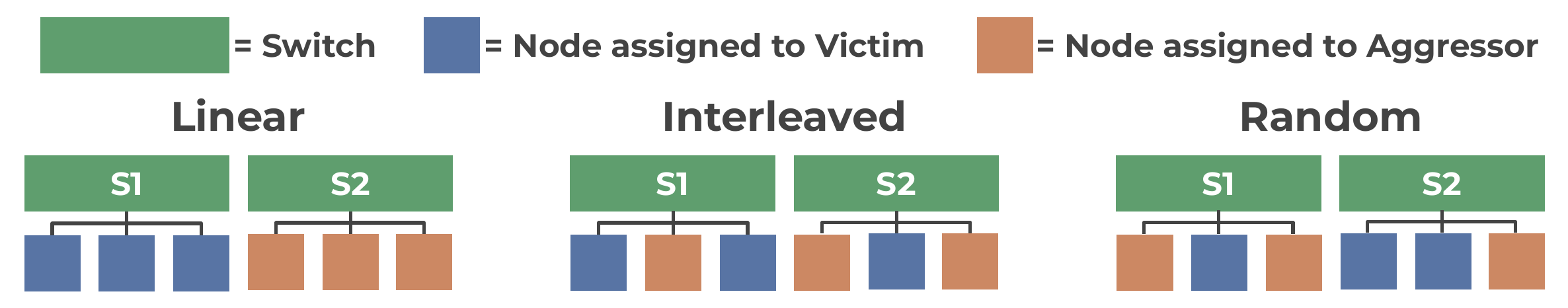}
    \caption{Different victim/aggressor allocations.}
    \label{fig:allocations}
\end{figure}

We make sure that the data we report is statistically sound~\cite{liblsb}:
for each microbenchmark, we execute the victim at least 200 times and for at least 4 seconds. We stop the benchmark when both the previous two conditions are satisfied, and when the 95\% confidence interval is within 5\% of the median. We then consider for each iteration the maximum time among the ranks. For the applications, we consider the time reported by the application, that we execute multiple times until the 95\% confidence interval is within 5\% of the median. 

We report in Figure~\ref{fig:evaluation:congestion:tailbench} the time distribution for the \textit{Tailbench} applications, both when executed in isolation, and when executed with an \textit{incast} aggressor, on both \aries and \slingshot. We also annotate the $99th$ and $95th$ percentiles, to show the impact of tail latency. We executed these experiments using the linear allocation and a $10\%/90\%$ victim/aggressor ratio. For \textit{Silo}, \textit{Xapian} and \textit{Img-dnn} we observe severe performance degradation due to congestion on \aries, whereas we do not observe any relevant effect on \slingshot. For \textit{Sphinx}, we observe a smaller degradation because the communication to computation ratio is lower than that of the other applications. Moreover, we observe a higher tail latency on \aries, which further increases in the presence of congestion.
It is worth remarking that the congestion impact itself is enough to characterize how much \slingshot is affected by congestion. In addition, we are also comparing \slingshot with an \aries interconnect, to also show the improvements compared to an existing interconnection network. Moreover, a similar performance degradation to that we observed on \aries has also been observed on other interconnects~\cite{gpcnet,fattree:sc18,htornnoise}.

\begin{figure}[htpb]
  \centering
  \includegraphics[width=\linewidth]{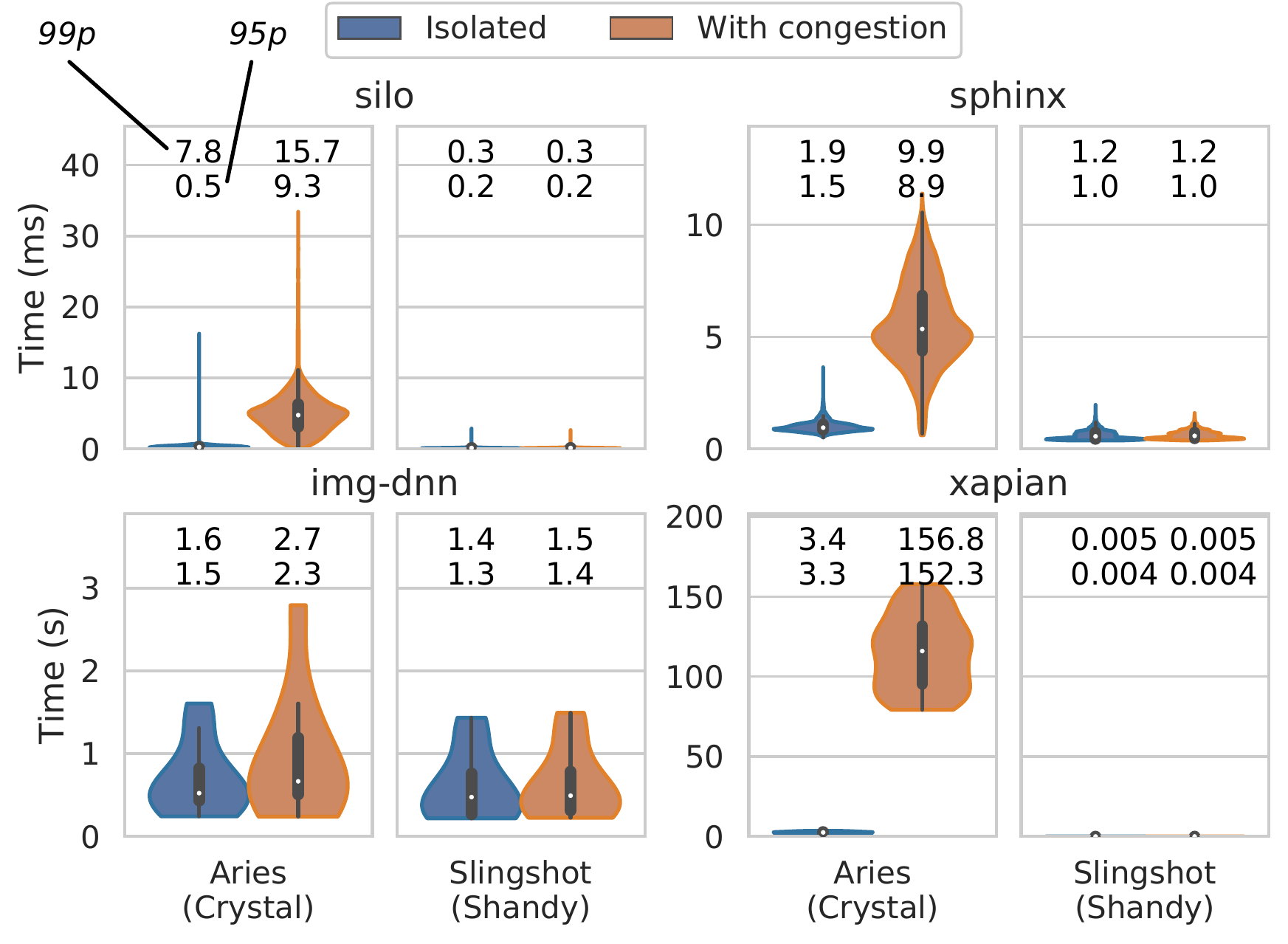}
\caption{Time distribution of \textit{Tailbench} applications, with and without endpoint congestion. The labels on the top of each plot denote the $99th$ and $95th$ percentile.}
\label{fig:evaluation:congestion:tailbench}
\end{figure}

Due to the large number of combinations of victims, aggressors, and allocations, we provide a data summary of the linear allocation results as a heatmap in Figure~\ref{fig:evaluation:congestion:heatmap}. Each element of the heatmap represents the mean congestion impact $C$~\cite{gpcnet}, i.e.,
\begin{equation}\label{eq:congimp}
C = \frac{T_{c}}{T_{i}}
\end{equation}

\begin{figure*}[h]
    \centering
    \includegraphics[width=\linewidth]{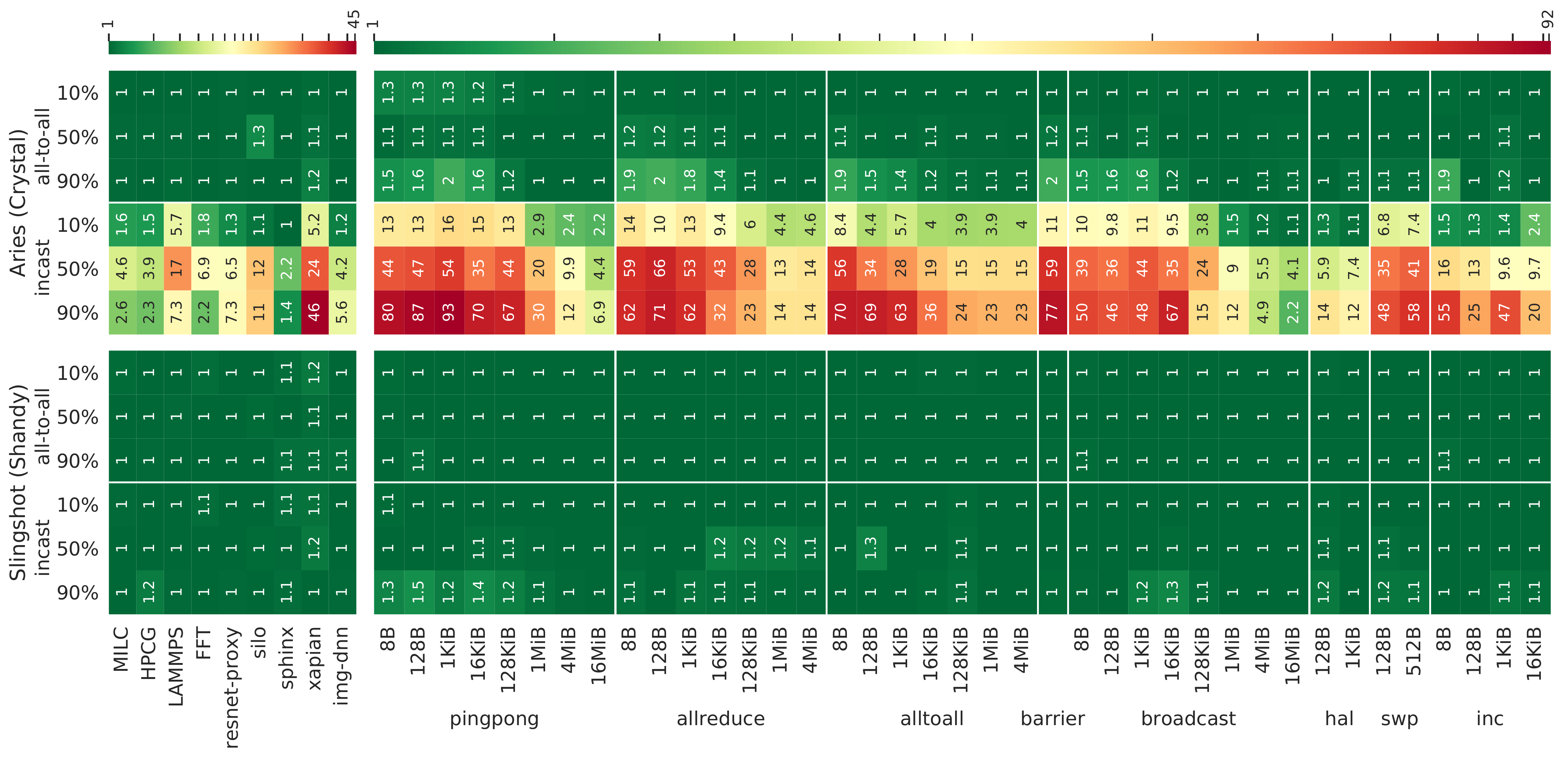}
    \caption{Congestion effects on different victim and aggressor combinations. Each element of the heatmap represents the congestion impact of the aggressor on the victim.}
\label{fig:evaluation:congestion:heatmap}
\end{figure*}

where $T_{i}$ is the mean execution time of the victim when executed in isolation, and $T_{c}$ is the mean execution time of the victim when co-executed with the aggressor. 
For example, the element on the top left corner represents the scenario where MILC is executed together with an \textit{all-to-all} aggressor. 10\% of the nodes are allocated to the aggressor, whereas the remaining nodes are allocated to the victim. For this specific case, no significant congestion impact is observed. On the other hand, MILC experiences a 1.6 slowdown on \aries due to endpoint congestion (\textit{incast}), when 10\% of the nodes are allocated to the aggressor. For the same scenario, we don't observe any slowdown on \slingshot.

We report the applications and microbenchmarks results using two different (logarithmic) color scales, to better appreciate the differences. Indeed, applications are usually less affected by congestion because, differently from microbenchmarks, they also have computation phases. Because communications are just a part of the overall execution time, even when communications are severely affected by congestion, this does not directly translate into a large performance degradation.

\begin{figure*}
\begin{minipage}{.75\linewidth}
\centering
\includegraphics[height=4.8cm]{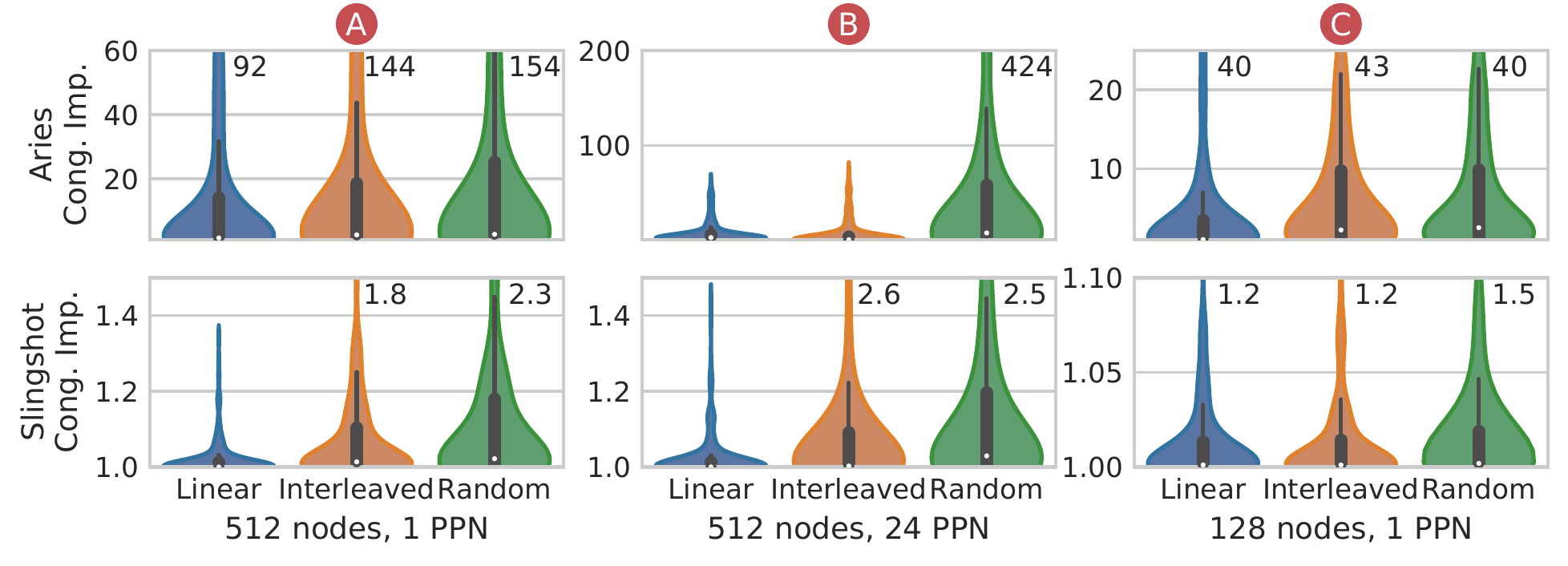}
    \captionof{figure}{Congestion impact distribution across different victim/aggressor combinations, for different allocations, node count, and processes per node (PPN).}
    \label{fig:evaluation:congestion:summaries}
\end{minipage}%
\hfill
\begin{minipage}{.23\linewidth}
  \centering
    \centering
    \includegraphics[height=4.8cm]{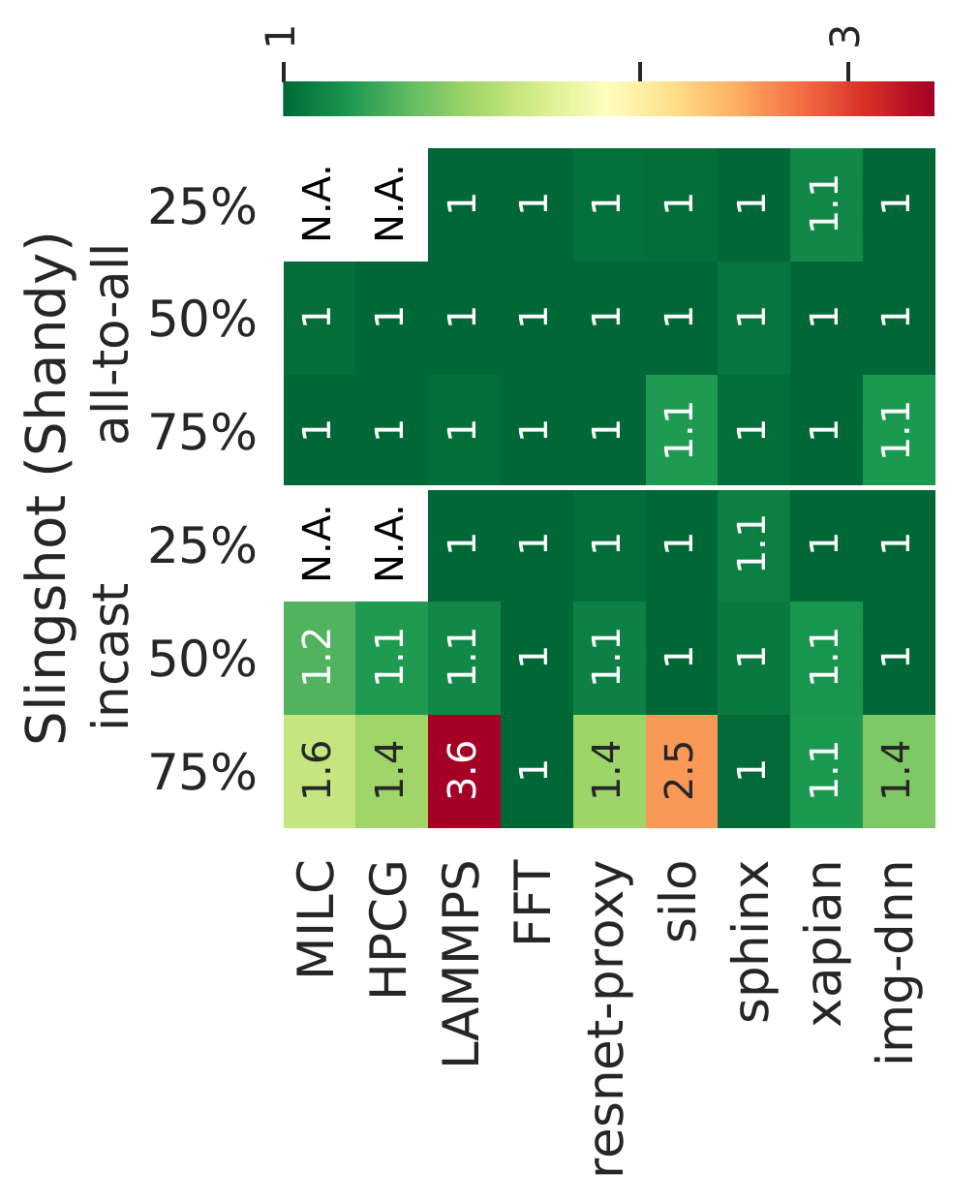}
    \captionof{figure}{Congestion impact on $1\,024$ nodes of \shandy.}
\label{fig:evaluation:congestion:1024}
\end{minipage}
\end{figure*}

As reported in the heatmap, \slingshot is always less affected by congestion compared to \aries. In the worst case, we observed a 1.3x slowdown on \slingshot, compared to a maximum 93x slowdown on \aries. Moreover, the congestion impact increases when increasing the fraction of nodes allocated to the aggressor application, and has a larger impact on small message communications, due to the larger impact of end-to-end latency on the overall performance. 
The effects of congestion can be seen not only on microbenchmarks but also on full applications. 
%
%
For example, LAMMPS is 17x slower when executed together with an \textit{incast} aggressor with a 50/50 split on \aries. Intermediate congestion (generated through \textit{all-to-all} communication), does not significantly affect the systems we are analyzing, because the adaptive routing algorithm successfully routes the packets around the congested links. This means that, 
the additional load generated by the \textit{all-to-all} does not manifest as congestion.

Similar trends can be observed also for different node count, higher PPNs, and other allocations. For space reasons, we do not report all the heatmaps for each of these cases. Instead, we summarize each heatmap by showing the distribution of the heatmap elements (congestion impacts, Equation~\ref{eq:congimp}) across all the victim/aggressor combinations. We show the result of this comparison in Figure~\ref{fig:evaluation:congestion:summaries}.

First, we show in Figure~\ref{fig:evaluation:congestion:summaries} (\includegraphics[scale=0.5,trim=0 3 0 0]{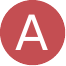}) the congestion impact for different allocations. For example, for the linear allocation, we are showing the same data of  Figure~\ref{fig:evaluation:congestion:heatmap}. However, instead of showing all the individual congestion impacts, we now report their distribution. For readability purposes, we cut the long tails of the distributions, and we annotate on top of each violin the maximum value. We observe that whereas on \aries the congestion impact for the linear allocation is never higher than 100, for the interleaved and random allocations we observe values up to 150. We observed a similar effect on \slingshot but on a different scale. In this case, in all but one cases we observe congestion impact values lower than two. Moreover, differently from \aries, the distribution on \slingshot is less spread, which indicates that the congestion control algorithm is performing well across a wide set of victims and allocations.

In Figure~\ref{fig:evaluation:congestion:summaries} (\includegraphics[scale=0.5,trim=0 3 0 0]{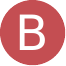}), we report a similar analysis, but now the aggressors are using 24 processes per node (PPN) instead of 1, thus generating a higher load on the network. In this case, the impact of congestion increases for \aries, especially for random allocations. On the other hand, 
\slingshot is only minimally impacted, showing a  maximum congestion impact $\sim 200$ times lower than on \aries.  


Lastly, in Figure~\ref{fig:evaluation:congestion:summaries} (\includegraphics[scale=0.5,trim=0 3 0 0]{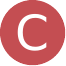}) we report the congestion impact when using fewer nodes (128). 
Until now, we compared a \slingshot system 
(\shandy, 1024 nodes) against a smaller \aries one (\crystal, 698 nodes). 
To factor out possible performance variations coming from different system sizes, 
we now compare \crystal with a smaller \slingshot system (\malbec, 484 nodes). 
We also fix the number of nodes per dragonfly group to 64, in order
to allocate the same number of groups (two) in both cases.
On \aries, the maximum congestion impact goes from 154 (Figure~\ref{fig:evaluation:congestion:summaries} (\includegraphics[scale=0.5,trim=0 3 0 0]{plots/contention/marker_a.pdf})) to 40 (Figure~\ref{fig:evaluation:congestion:summaries} (\includegraphics[scale=0.5,trim=0 3 0 0]{plots/contention/marker_c.pdf})) when using 128 nodes instead of 512. This can be explained
by the lower generated traffic (aggressors have now fewer nodes), but also by
the higher fraction of available global bandwidth. 
On \slingshot, the same experiment makes the maximum congestion impact go
from 2.3 to 1.5. 
We conclude that \slingshot is less affected by congestion, even when varying
the system size and the number of allocated nodes.

The results of Figure~\ref{fig:evaluation:congestion:1024} show the congestion impact on the applications when using all the $1\,024$ nodes on \shandy. We report the data when using a random allocation because that is the one generating the most congestion (see Figure~\ref{fig:evaluation:congestion:summaries}). We can observe that even at full system scale the congestion control effectively protects applications from congestion, with a maximum 3.55x slowdown on LAMMPS when 75\% of the nodes are allocated to the \textit{incast} congestor. Data on MILC and HPCG with a 25\%/75\% aggressor/victim ratio is missing. Indeed, they should run on 768 nodes, but they can only run on a number of nodes which is a power of two.

We complete our analysis on the effects of congestion by analyzing the impact of bursty congestion \slingshot. Indeed, in the previous experiments we always considered persistent congestion, generated by sending messages with a fixed size of 128KiB during the entire victim execution. To analyze the impact of bursty congestion, we execute a 128 byte \texttt{MPI\_Alltoall} microbenchmark (victim) with an \textit{incast} aggressor. This is one of the cases where we observed the highest congestion impact on \slingshot (see Figure~\ref{fig:evaluation:congestion:heatmap}). We run this test on all the \malbec nodes, splitting them equally between aggressor and victim, with an interleaved allocation strategy.

\begin{figure}[h]
    \centering
    \includegraphics[width=\columnwidth]{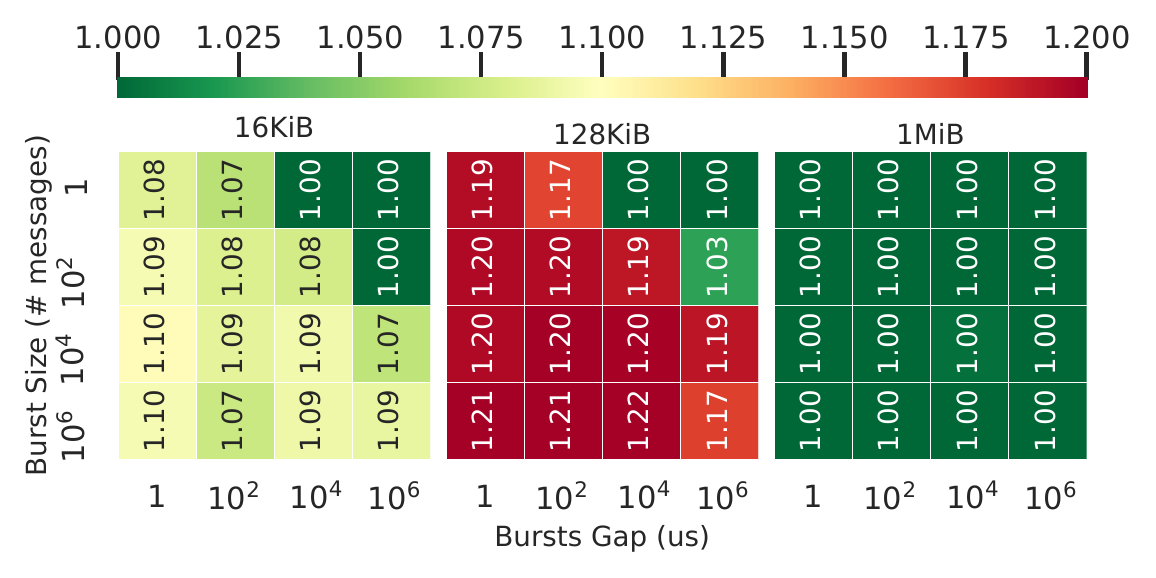}
    \caption{Impact of \textit{incast} congestion on a 128 byte \texttt{MPI\_Alltoall}. We show the impact for different message sizes, congestion duration, and time between subsequent congestion bursts.}
    \label{fig:congestion_parameters}
\end{figure}

We report the results of this analysis in Figure~\ref{fig:congestion_parameters}. Each heatmap corresponds to a different message size for the \textit{incast} aggressor. On each heatmap we report the congestion impact when varying the number of messages in a burst (\textit{Burst Size}, on the y-axis) and the time between two subsequent congestion bursts (\textit{Bursts Gap}, on the x-axis). For example, the bottom-left element in the first heatmap, represents the case where the aggressor sends $10^6$ consecutive messages, each one containing 8 bytes. Before sending the next burst of $10^6$ messages, the aggressor will wait 1 microsecond.

We observe that the \textit{incast} aggressor does not affect the victim when sending too small messages or too large messages. Indeed, small messages do not generate enough congestion, whereas for large messages the congestion control algorithm fully kicks in and throttle the aggressor. On the other hand, for medium size messages, some congestion builds up before the congestion control algorithm detects and reacts to it, and we observe an increase in the congestion impact up to $1.21$. However, as we shown in Figure~\ref{fig:evaluation:congestion:heatmap}, this is negligible when compared to what happens on other types of systems. Moreover, we observe the highest congestion impact for large bursts and for small gaps between subsequent bursts. 
We also observe no differences between bursts of $10^6$ messages and the persistent congestion. This shows that \slingshot is tolerant to both persistent congestion, and bursty and short-lived congestion.

\subsection{Traffic Classes}\label{sec:evaluation:classes}
We now evaluate the ability of \slingshot to provide performance guarantees to jobs running by using traffic classes. It is worth remarking that traffic classes and congestion control are orthogonal concepts. Traffic classes can be used to protect a job (or parts of it) from other traffic, and they can allocate resources fairly or unfairly between users and jobs. However, even if resources are assigned fairly, congestion can still occur due to jobs filling up the buffers. Congestion control is used to avoid such situations within and across traffic classes.

All the experiments presented in the following have been executed on \malbec. We taper the bandwidth to 25\% of the available bandwidth, to force co-running jobs to interfere with each other. We execute a job performing an 8B \texttt{MPI\_Allreduce} together with a job performing a 256KiB \texttt{MPI\_Alltoall}. Each job uses 64 nodes and 16 processes per node, and they are placed using the interleaved allocation. We report in Figure~\ref{fig:qos:ara2a} the congestion impact of the \texttt{MPI\_Allreduce} when using the same traffic class of the \texttt{MPI\_Alltoall} and when using a separate traffic class. Each point represents the mean over $100\,000$ runs. The \texttt{MPI\_Alltoall} is started around 0.4 milliseconds after the beginning of the test. We observe that when \texttt{MPI\_Allreduce} runs in the same traffic class of the \texttt{MPI\_Alltoall}, it experiences a congestion impact of 2.85 (i.e. is 2.85 times slower compared to when executed in isolation). On the other hand, when executed in a separate traffic class it only experiences a 1.15x slowdown compared to the isolated case. 

We now further investigate the capacity of \slingshot to enforce specific limits on traffic classes. We execute two jobs, each running a bisection bandwidth test, with the second one starting after 0.9 milliseconds from the beginning of the test. Each job uses 16 processes per node and runs on 64 nodes. Jobs are placed by using the interleaved allocation. We configure two traffic classes: \texttt{TC1} with a minimum bandwidth requirement of 80\% of the available bandwidth, and \texttt{TC2}, with a minimum 10\% bandwidth required. 

\begin{figure}
  \centering
  \includegraphics[width=\linewidth]{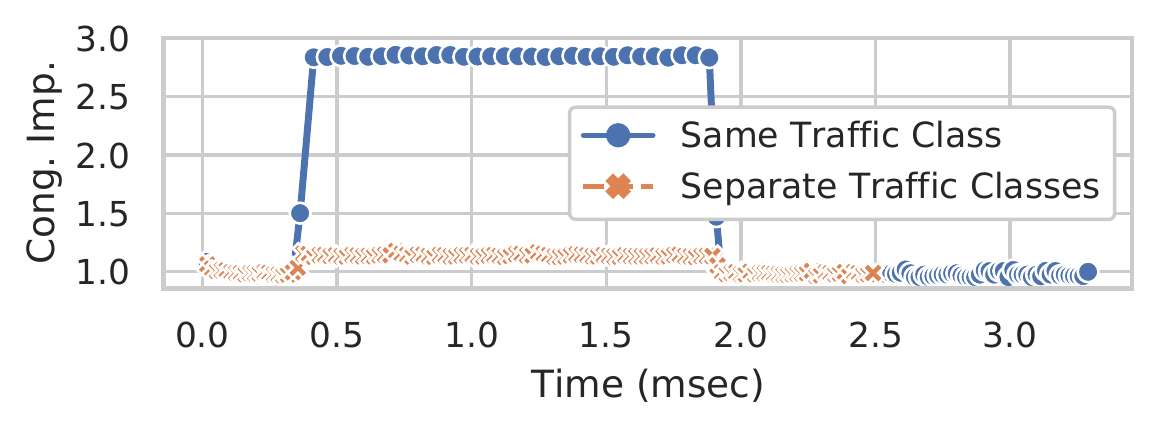}
    \caption{Congestion impact for an 8B \texttt{MPI\_Allreduce}, co-executed with a 256KiB \texttt{MPI\_Alltoall} on \malbec (with a 25\% tapering) with and without traffic classes.}
    \label{fig:qos:ara2a}
\end{figure}

We report the results of this experiment in Figure~\ref{fig:qos:bisect}. On the upper part, we report the results we obtain when both jobs run on the same traffic class (\texttt{TC1}). At the beginning of the execution, the first job runs on an empty system and gets 100\% of the available bandwidth. When the second job starts, the available bandwidth is fairly shared between the two jobs. Eventually, when the first job terminates the second job ramps up and uses all the available bandwidth. 

\begin{figure}[h]
  \centering
  \includegraphics[width=\linewidth]{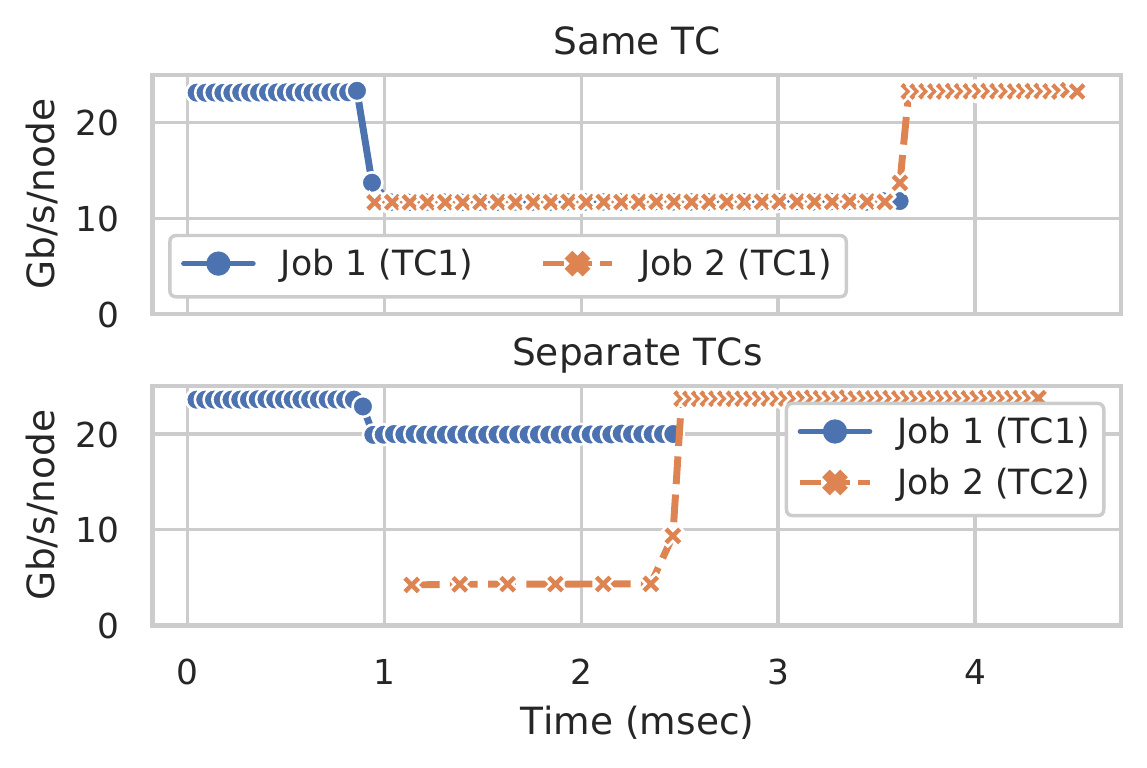}
  \caption{Performance of two bisection bandwidth tests on \malbec (with a 25\% tapering) when running in the same traffic class (top) and when running into two separate traffic classes (bottom).}
  \label{fig:qos:bisect}
\end{figure}

On the lower part of Figure~\ref{fig:qos:bisect}, we report the results when the first job runs in \texttt{TC1} and the second job runs in \texttt{TC2}. In this case, when the second job starts, the bandwidth of the first job drops to 80\% of the available bandwidth, matching the minimum bandwidth required for \texttt{TC1}. The second job required a minimum bandwidth of 10\%, and it gets the 20\% of the available bandwidth. Indeed, there is an extra 10\% of bandwidth which was not allocated to either \texttt{TC1} or \texttt{TC2}. \slingshot decides to dynamically allocate this extra bandwidth to \texttt{TC2} because it is the traffic class with the lowest bandwidth share. Eventually, when the first job terminates, the second job uses all the available bandwidth.

\section{State of the Art}\label{sec:state}
\subsection{Interconnection Networks}
Existing large-scale computing systems are characterized by different types of interconnection networks, either based on open standards or proprietary technology. These networks have different topologies and provide different features. In this section, we highlight the main characteristics of the most common and actively developed interconnection networks, to better understand the similarities and differences with \slingshot.

\textit{InfiniBand} is an open standard for high-performance network communications. Different vendors manufacture InfiniBand switches and interfaces, and the InfiniBand standard is not tied to any specific network topology. The most commonly used InfiniBand implementations rely on \textit{Mellanox} hardware, with switches arranged in a \textit{fat tree} topology~\cite{fattree}. For example, both \textit{Sierra}~\cite{sierra} and \textit{Summit}~\cite{summit}, the two fastest supercomputers at the time being, use such configuration. 
Mellanox networks also provide other features to improve application performance, such as switch offloading of MPI collective operations, adaptive routing, congestion control, and traffic classes. However, congestion control is usually not used in large production systems due to difficulties in the tuning of the algorithm~\cite{gpcnet}. Regarding interoperability with Ethernet, Mellanox adopts a different approach than \slingshot, requiring traffic to be converted between InfiniBand and Ethernet by using dedicated gateways.

Cray \aries~\cite{crayxcpdf} is the 7th generation of Cray interconnection networks. It is based on a Dragonfly topology and supports different systems configuration up to $92\,544$ nodes (Trinity~\cite{trinity}, the largest \aries system currently deployed, has $19\,420$ nodes). It provides a peak injection bandwidth of 81.6 Gb/s per node, and a rich set of features including adaptive routing, collective operations offload, and remote atomic operations. It uses fewer optical links than \textit{fat trees} networks, reducing the cost of the network. 

\textit{Tofu Interconnect D} (TofuD)~\cite{tofud} is the third generation \textit{Tofu} interconnection networks, which will be used by the \textit{Fugaku} supercomputer~\cite{fugaku} (formerly known as \textit{Post-K}). \textit{TofuD} provides a peak injection rate of 300Gb/s per node and, like its predecessors, it is based on a 6D mesh/torus. Around 25\% of the links used by the interconnect are optical. To reduce latency and improve fault resiliency,  \textit{TofuD} uses a technique called \textit{dynamic packet slicing}, to split the packets in the data-link layer. This can either be used to split the packet and improve the transmission performance or to duplicate the packet to provide fault tolerance in case the link quality degrades. Moreover, this interconnect provides an offload engine, called \textit{Tofu Barrier}, to execute collective operations without involving the CPU.

The \textit{Dragonfly+}~\cite{dragonflyplus} is currently used by the \textit{Niagara} supercomputer~\cite{niagara}. It is a variation of the Dragonfly interconnect~\cite{dragonfly}, where the switches inside a group are connected through a fat-tree network. Similarly to the Dragonfly network, this interconnect is characterized by different minimal and non-minimal paths between each pair of nodes. The implementation used in the Niagara supercomputer relies on Mellanox InfiniBand hardware. To select the optimal path, \textit{Dragonfly+} uses a variation of the \textit{OFAR} adaptive routing~\cite{ofar}, which at each hop re-evaluates the optimal path to use. Explicit control messages are sent among the switches to notify congestion and avoid creating hotspots in the network. 


Several other low-diameter networks~\cite{diam2paper} have been proposed by the research community, including but not limited to \textit{SlimFly}~\cite{slimfly},  \textit{Megafly}~\cite{megafly},  \textit{HyperX}~\cite{hyperx,hyperx-implementation},  \textit{Jellyfish}~\cite{jellyfish} and \textit{Xpander}~\cite{xpander} topologies.
On the data centers side, Clos~\cite{clos1953study} is the most prevalent deployed topology. Whereas the above mentioned low-diameter topologies are claiming to have substantial cost-performance improvements, they have been scarcely employed because of hard-to-deploy routing schemes. Also, classical congestion control mechanisms (e.g., ECMP~\cite{hopps2000analysis}) are not effective in such low-diameter networks due to the scarcity of minimal paths~\cite{fatpaths}. \slingshot addresses these issues by providing a 
low-diameter network with an effective congestion control algorithm, setting a stepping stone towards HPC data centers.

Overall, \slingshot introduces a set of key features that can be taken as reference for next-generation large-scale computing systems. 
First, the end-to-end congestion control algorithm can quickly react to congestion and is stable across a wide set of applications and microbenchmarks. Moreover, traffic classes provide additional flexibility and open new software optimization opportunities. Lastly, it is natively interoperable with existing Ethernet devices, and thanks to novel adaptive routing strategies, it provides high network utilization also for in-order RoCE traffic (see Figure~\ref{fig:alltoall}).


\subsection{Interconnection Networks Benchmarking}
In this work we described the \slingshot interconnection network and, for the first time, we extensively evaluated it across a wide set of microbenchmarks and real applications. We reported both the isolated performance and the performance under the presence of congestion. 

Regarding the evaluation of the under-load system, different works analyzed the impact of congestion (also known as \textit{network noise}) on application performance~\cite{appawarerouting,gpcnet,10.1145/3295500.3356168,10.1109/SC.2018.00030,10.5555/3014904.3014990,htornnoise,6877474} on different types of networks. The GPCNet benchmark~\cite{gpcnet} has been recently proposed as a portable benchmark for estimating network congestion. We used in this work the same definition of endpoint/intermediate congestion and of \textit{congestion impact} used by GPCNet.
Whereas the authors of GPCNet also report some preliminary results on a \slingshot system, they do not provide a detailed view of the system performance. Indeed, the main goal of GPCNet was to design a portable congestion benchmarking infrastructure by using a small set of victim microbenchmarks (random ring and \texttt{MPI\_Allreduce}) to easily compare different systems. However, this does not represent a wide spectrum of real scenarios.
On the other hand, we focus on the impact of congestion on \slingshot by using different microbenchmarks and both on HPC and datacenters applications. Moreover, the GPCNet paper only analyzes the impact of congestion for a fixed victim message size, allocation, and aggressor/victim ratio. However, as we show in Section~\ref{sec:evaluation:congestion}, all these factors play a role in the observed congestion and they can be helpful to understand the system performance. 

\section{Conclusions}\label{sec:conclusions}
Interconnection networks have a significant impact on the performance of large computing systems, both in supercomputers and hyperscale datacenters. In this paper, we describe and evaluate \slingshot, the latest interconnection network designed by Cray. We describe \slingshot's main features: high-radix Ethernet switches, adaptive routing, congestion control, and QoS management. We then evaluate \slingshot's performance, both in isolation and when executing different concurrent workloads. 

Our results demonstrate that applications running on \slingshot are much less affected by congestion compared to previous generation networks and that the congestion control algorithm works on a wide set of different microbenchmarks and HPC and datacenter applications. We also show that allocation policies have a much lower impact on performance on \slingshot compared to previous generation networks.  Lastly, we demonstrate how \slingshot can provide bandwidth guarantees to jobs running in separate traffic classes. 

The information we provide can be used by HPC and datacenter system operators, administrators, users, and programmers to optimize, deploy, and manage parallel applications. A deep understanding of the interconnect's features is a prerequisite to ensure optimized operations and utilization of computing resources in clouds and datacenters. 

\section*{Acknowledgment}
We thank the anonymous reviewers for their insightful comments, and the Slingshot team at HPE for providing access and support in using the systems. We thank Steve Scott (HPE) for invaluable input. We would also like to thank Shigang Li for providing the code for the \emph{Resnet-proxy} application. Daniele De Sensi is supported by an ETH Postdoctoral Fellowship (19-2 FEL-50). This project has received funding from the European Research Council (ERC) under the European Union’s Horizon 2020 programme (grant agreement DAPP, No. 678880). 

\bibliographystyle{unsrt}
\bibliography{bibfile}

\end{document}